\documentclass[epj,twocolumn]{webofc}
\usepackage[varg]{txfonts}   % Web of Conferences font

\usepackage{notoccite}
\usepackage{amsmath}
\usepackage{graphicx}
\usepackage{color}

\begin{document}

\title{Measurement of the charged pion mass using a low-density target of light atoms}
%
% subtitle is optionnal
%%%\subtitle{Do you have a subtitle?\\ If so, write it here}

\author{\firstname{M.} \lastname{Trassinelli}\inst{1}\fnsep\thanks{\email{martino.trassinelli@insp.jussieu.fr}} 
	\and \firstname{D.F.} \lastname{Anagnostopoulos}\inst{2}
	\and \firstname{G.} \lastname{Borchert}\inst{3}\fnsep\thanks{present address: TU Munich, D-85747 Garching, Germany}
	\and \firstname{A.} \lastname{Dax}\inst{4}
	\and \firstname{J.-P.} \lastname{Egger}\inst{5}
	\and \firstname{D.} \lastname{Gotta}\inst{3}\thanks{\email{d.gotta@fz-juelich.de}} 
	\and \firstname{M.} \lastname{Hennebach}\inst{3}\fnsep\thanks{present address: DAHER NUCLEAR TECHNOLOGIES GmbH, D-63457 Hanau, Germany}
	\and \firstname{P.} \lastname{Indelicato}\inst{6}
	\and \firstname{Y.-W.} \lastname{Liu}\inst{4}\fnsep\thanks{present address: Phys. Depart., National Tsing Hua Univ., Hsinchu 300, Taiwan}
	\and \firstname{B.} \lastname{Manil}\inst{6}\fnsep\thanks{present address: Lab. de Physique des Lasers, Universit\'e Paris 13, Sorbonne Paris Cit\'e, CNRS, France}
	\and \firstname{N.} \lastname{Nelms}\inst{7}\fnsep\thanks{present address: ESA-ESTEC, PO Box 299, 2200 AG, Noordwijk, The Netherlands}
	\and \firstname{L.M.} \lastname{Simons}\inst{4}
	\and \firstname{A.} \lastname{Wells}\inst{7}
        % etc.
}

\institute{Institut des NanoSciences de Paris, CNRS, Sorbonne Universit\'es, UPMC Univ Paris 06, F-75005, Paris, France 
\and
           Dept. of Materials Science and Engineering, University of Ioannina, GR-45110 Ioannina, Greece
\and
           Institut f\"ur Kernphysik, Forschungszentrum J\"ulich GmbH, D-52425 J\"ulich, Germany
\and
           Laboratory for Particle Physics, Paul Scherrer Institut, CH 5232-Villigen PSI, Switzerland
\and
           Institut de Physique de l' Universit\'{e} de Neuch\^{a}tel, CH-2000 Neuch\^{a}tel, Switzerland
\and
           Laboratoire Kastler Brossel, UPMC-Sorbonne Universités, CNRS, École Normale Supérieure-PSL Research University, Collège de France, F-75005 Paris, France
\and
           Dept. of Physics and Astronomy, University of Leicester, Leicester LEI7RH, England
          }

\abstract{%
We present a new evaluation of the negatively charged pion mass based on the simultaneous spectroscopy of pionic nitrogen and muonic oxygen transitions using a gaseous target composed by a N$_2$/O$_2$ mixture at 1.4~bar.
We present the experimental set-up and the methods for deriving the pion mass value from the spatial separation from the $5g-4f$ $\pi$N transition line and the $5g-4f$ $\mu$O transition line used as reference.
Moreover, we discuss the importance to use dilute targets in order to minimize the influence of additional spectral lines from the presence of remaining electrons during the radiative emission. 
The occurrence of possible satellite lines is investigated via hypothesis testing methods using the Bayes factor.

}
\maketitle
\section{Introduction}
\label{intro}

\begin{figure}
% Use the relevant command for your figure-insertion program
% to insert the figure file.
\centering
\includegraphics[width=\columnwidth,clip,trim={40 30 15 30}]{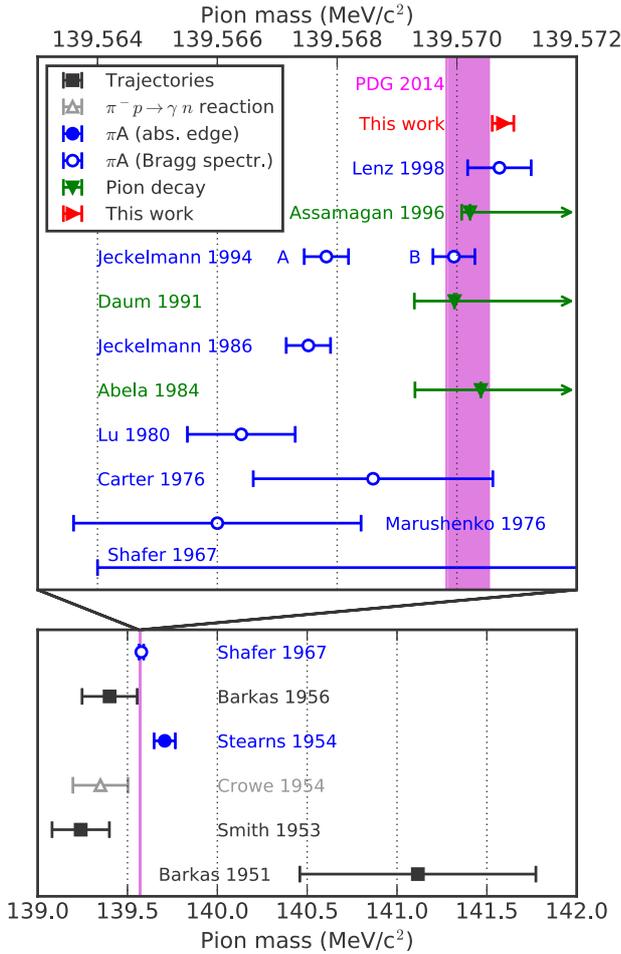}
\caption{Results of for the mass of the charged pion from various experimental methods since its discovery. The world average value ``PDG 2014'' \cite{PDG2014}, indicated by the magenta region, is calculated from ``Lenz 1998'' \cite{Lenz1998} and solution B of ``Jeckelmann 1994'' \cite{Jeckelmann1986b}. Other results are taken from Refs.~\citenum{Jeckelmann1986,Jeckelmann1986b,Daum1991,Abela1984,Lu1980,Carter1976,Marushenko1976,Shafer1967,Barkas1956,Stearns1954,Crowe1954,Smith1953,Barkas1951}.}
%\citenum{Jeckelmann1986,Jeckelmann1986b,Daum1991,Abela1984,Lu1980,Carter1976,Marushenko1976,Shafer1967,Barkas1956,Stearns1954,Crowe1954,Smith1953,Barkas1951}.}
%``Daum 1991'' \cite{Daum1991}, ``Jeckelmann 1986''\cite{Jeckelmann1986,Jeckelmann1986b}, ``Abela 1984'' \cite{Abela1984}, ``Lu 1980'' \cite{Lu1980}, ``Carter 1976'' \cite{Carter1976}, ``Marushenko 1976'' \cite{Marushenko1976}, ``Shafer 1967'' \cite{Shafer1967}, ``Barkas 1956'' \cite{Barkas1956}, ``Stearns 1954'' \cite{Stearns1954}, ``Crowe 1954'' \cite{Crowe1954}, ``Smith 1953'' \cite{Smith1953} and ``Barkas 1951'' \cite{Barkas1951}.}
\label{fig:history}       % Give a unique label
\end{figure}

The first estimation of the charged pion mass came with its discovery from cosmic rays traces in photographic plates \cite{Lattes1947a,Lattes1947b}.
Counting the photographic emulsion grains in the particle trajectory, Powell and his group estimated the pion-to-muon  mass ratio to be about 1.5.
With the pion production from accelerators \cite{Gardner1948}, the charged pion mass has been measured from the deflection trajectory in a magnetic field \cite{Barkas1951,Smith1953,Barkas1956} but also from the energy of the gamma-ray produced by the reaction $\pi^- + p \to n +\gamma$ \cite{Crowe1954} (see Fig.~\ref{fig:history}).
The first mass measurement employing pionic atoms was performed in 1954 by using $4f \to 3d$ X-ray emission of several light pionic atoms with the critical absorption edge technique  reaching an accuracy of 0.5\% \cite{Stearns1954}.
With the increase of intensity of pion beams, the use of crystal spectrometers has become possible. In 1967, the spectroscopic measurement of pionic calcium and titanium  reached the relative accuracy of $10^{-4}$ \cite{Shafer1967}.
As shown in Fig.~\ref{fig:history}, in the following years the pion mass has been re-measured several times using X-ray spectroscopy of pionic atoms with continuously improved accuracy \cite{Marushenko1976,Carter1976,Lu1980,Jeckelmann1986,Jeckelmann1986b,Jeckelmann1994,Lenz1998}.
In parallel, additional results for the charged pion mass were obtained from the measurement of the muon momentum in the decay $\pi^+ \to \mu^+ \nu_\mu$ \cite{Abela1984,Daum1991,Assamagan1996}.

With increasing experimental accuracy, in the '80s  discrepancies between pionic atoms and pion decay results showed up. 
The disagreement was due to the ambiguous assumption on the remaining electrons in pionic atoms when produced in solid-state targets. 
In the $\pi$Mg experiment of Jeckelmann and collaborators \cite{Jeckelmann1986,Jeckelmann1986b}, where the $4f-3d$ transition (25.9~keV) was measured with a transmission X-ray DuMond spectrometer, different assumptions on the K electron population lead to a difference in the pion mass of 16~ppm between the two possible interpretations, called A and B\,\cite{Jeckelmann1994}. 
In particular, solution A is in complete disagreement with evaluations obtained from the pion decay measurement at rest  \cite{Abela1984,Daum1991,Assamagan1996}.

To solve this dilemma, in the '90s our collaboration designed and realized a new experiment with a gaseous nitrogen target for having X-rays emitted from a purely hydrogen-like pion-nucleus system.
In this experiment, the energy of the 4~keV X rays of the $5g \to 4f$ line was measured with a reflection spectrometer set up in Johann geometry. The spectrometer was previously developed for light antiprotonic atom spectroscopy \cite{Gotta1999}. 
This experiment's limitation was the calibration line, the Cu~K$\alpha$ fluorescence radiation, whose large natural width (4~eV) prevents an energy determination at ultimate precision.

The present pion-mass value given by the Particle Data Group (PDG) \cite{PDG2014} has an accuracy of 2.5~ppm and is the result of the average of solution B of the measurement of pionic magnesium \cite{Jeckelmann1994}  and the one obtained from  pionic nitrogen spectroscopy \cite{Lenz1998}.

\begin{figure}
% Use the relevant command for your figure-insertion program
% to insert the figure file.
\centering
\includegraphics[width=0.9\columnwidth,clip]{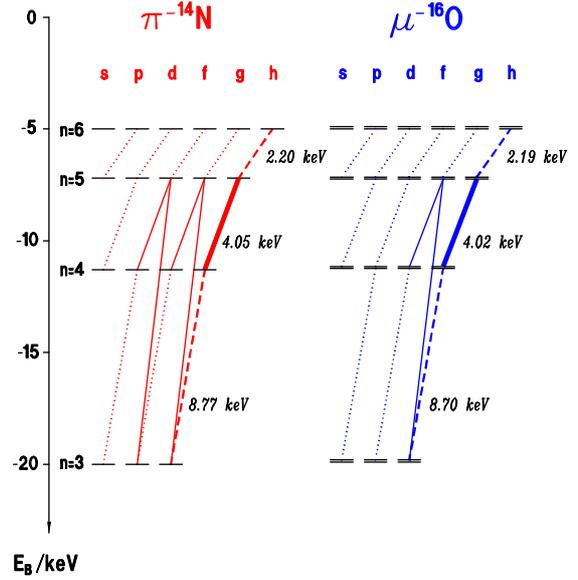}
\caption{Comparison of the intermediate parts of the cascade  for pionic nitrogen and muonic oxygen.}
\label{fig:cascade}       % Give a unique label
\end{figure}

The experiment described here resumes the strategy of gas targets, but exploits (i) the high precision of 0.033~ppm for the mass of the positively charged muon being m$_{\mu^{+}} = (105.6583715 \pm 0.0000035)$\,MeV/c$^2$\,\cite{PDG2014} and (ii) the unique feature that in $\pi$N and $\mu$O transition energies almost coincide (Fig.~\ref{fig:cascade}).  
When using a gas mixture, the simultaneous measurement of $\pi$N and $\mu$O lines becomes possible with the muonic transition serving as an on-line calibration. 

The experiment and some aspects of the data analysis have been described in Ref.~\citenum{Trassinelli2016b}. 
Here, we present additional aspects. In particular, we discuss in detail the possibility of remaining electrons during the radiative dominated part of the cascade  (Sec.~\ref{sec:cascade}) applying specific analyses and statistical tests to our data (Sec.~\ref{sec:satellite}).
We present a brief general description of the experiment (Sec.~\ref{sec:spectrometer}) and the detailed formulas used for extracting the pion mass from the line positions (Sec.~\ref{sec:formulas}).

\section{Pionic atoms production and atomic cascade}\label{sec:cascade}

\begin{figure}
% Use the relevant command for your figure-insertion program
% to insert the figure file.
\centering
\includegraphics[width=0.9\columnwidth,clip]{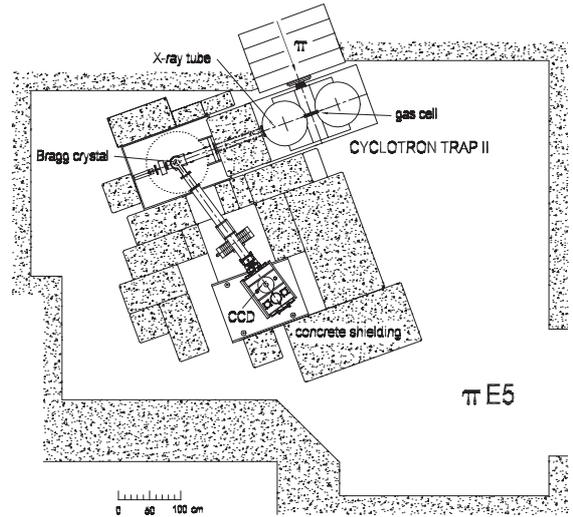}
\caption{Experimental set-up with at the $\pi$E5 beam line of the Paul Scherrer Institut (Villigen, Switzerland). The pions enter the cyclotron trap from top perpendicular to its magnetic field.}
\label{fig:setup}       % Give a unique label
\end{figure}

The experiment was performed by using the intense pion beam at the $\pi$E5 beam line of the Paul Scherrer Institut. 
A general overview of the set-up is presented in Fig.~\ref{fig:setup}.
The pions, with an initial momentum of 112~MeV/c, are captured and slowed down in the so-called cyclotron trap II.
The cyclotron trap consists of a superconducting split-coil magnet with a field perpendicular to the pion trajectories. In a set of plastic degraders, pions are slowed  down in order to be stopped in the gas contained in a target cell at center of the trap.
A first version of this trap had been initially developed to efficiently decelerate and trap antiprotons for antiprotonic atoms production \cite{Simons1988}.
Lenz \textit{et. al.} \cite{Lenz1998} used this trap to decelerate and stop pions.

The cyclotron trap II used in the present experiment was specifically designed with a larger gap between the magnet coils. This allows a more efficient trapping of muons produced in the decay of slow pions, which have to be captured in the bottle field of the magnet before being stopped in the target \cite{Simons1993}. 
$1-3\%$ of the initial pions are stopped in the gas inside the target cell. The stop rate of the decay muons is about 10\% of the one of the pions.
%The large gap between the magnet coils and the strong magnetic field (up to 4~T) guarantee an efficient recapture of muons and then an efficient stop rate for the production of muonic atoms with a lower efficiency.

For the simultaneous measurement of $5g-4f$ transitions from pionic nitrogen and muonic oxygen, comparable count rates are required. 
This was achieved with a N$_{2}$/O$_{2}$ mixture of 10\%/90\%. The gas was kept at a pressure of 1.4~bar and room temperature. 

Pions and muons are captured where the overlap of the wave-functions of the outermost electrons and of the pion or muon is largest. Electronic quantum numbers $n_e$ correspond to highly exited states with an initial quantum number $n_i \sim n_e \times 16 ~(14)$ for pions (muons), from where a quantum de-excitation cascade starts\,\cite{Cohen2004}.

For exotic atoms with atomic number $Z>2$, the formation and the first steps of the de-excitation proceed via Auger emission and self-ionization of the target atom or molecule.
Acting at the femtosecond time scale, this process quickly leads to a high degree of ionization.

Auger emission determines the upper part of the atomic de-excitation cascade where the radiative emission is dominating for lower levels. The Auger transition probability $\Gamma_A$ is proportional to $1/\sqrt{2 \Delta E + 37.8~\text{eV}}$, where $\Delta E$ is the transition energy, favouring $\Delta n=1$ transitions with the selection rule $\Delta \ell = 0, \pm 1$   \cite{Burbidge1953,Leon1962}.
The radiative transition probability $\Gamma_X$ is proportional to $\Delta E^3$ with selection rules $\Delta \ell = \pm 1$ and, hence, maximal $\Delta n$ de-excitation steps are preferred. Such a dependence on $\Delta E$ efficiently populates $| n, \ell=n-1 \rangle$ circular states allowing subsequently only slow  radiative $(n,\ell =n-1)\rightarrow (n-1,\ell =n-2)$ transitions after depletion of the electron shells. Therefore, the atomic cascade duration is essentially  determined by the first radiative transitions.

If the target consists of molecules, the initial ionisation causes a repulsive Coulomb force between the atomic cores leading to an acceleration (Coulomb explosion) where the captured pion or muon is attached to one of the atoms.
After Coulomb explosion, Auger emission continues until the  complete depletion of the electron shells if capture from neighbouring target atoms is avoided.

If a low-density target ($\le 1-2$~bar) is used, electron recapture from external atoms is unlikely because the probability for having a collision with another atom of the target is low even in the presence of Coulomb explosion. This is proven by the appearance of X-ray lines at $n\geq5$, which otherwise would be converted into Auger transitions \cite{Burbidge1953,Vogel1980,Bacher1985,Bacher1989,Kirch1999}. 
In the case of solid targets, however, electron refilling is unavoidable.

The circular transitions $(n,\ell = n-1) \to (n-1,\ell=n-2)$ are by far the most intense X-ray lines and, therefore, to be used for the measurement. For these transitions, strong interaction effects in the case of pionic atoms are minimal but  Coulomb explosion leads to a significant Doppler broadening of the X-ray lines \cite{Siems2000}. More details on the atomic cascade in exotic atoms can be found in Refs.~\cite{Jensen2002a,Jensen2002b,Cohen2004,Kilic2004,Pomerantsev2006,Jensen2007,Popov2007}.

In our experiment, electron recapture during the cascade is not expected when using a N$_{2}$/O$_{2}$ mixture with a pressure of 1.4~bar at room temperature. This hypothesis is  explicitly tested in the analysis presented here (Sec.~\ref{sec:satellite}).

\section{X-ray spectrometer and data acquisition} \label{sec:spectrometer}

The exotic atom transition energies are precisely measured using Bragg diffraction spectroscopy.
The crystal spectrometer is set up in Johann geometry \cite{Johann1931} and has been specifically designed for high-accuracy X-ray spectroscopy of light exotic atoms \cite{Gotta2004,Gotta2016} and was initially used for antiprotonic hydrogen and deuterium spectroscopy at LEAR \cite{Gotta1999}.
The Johann  configuration allows the simultaneous measurement of two different energies.
The limit of the acceptable energy difference is given by the extension of the X-ray source (target) and/or of the detector on the focus position. 

The Bragg crystal was made from a silicon crystal disk cut along the $220$ plane of 290\,$\mathrm{\mu}$m thickness with a diameter of 100\,mm. 
The disk is attached to a high-quality polished glass lens defining a spherical segment. Its curvature was determined to be $(2981.3\pm0.3)$\,mm. 
Spherical bending leads to a partial vertical focusing \cite{Eggs1965} which increases the count rate.

For the simultaneous detection of the pionic and muonic atoms lines, the spectrometer has been equipped with a large position-sensitive detector, which is composed of an array of charge-coupled devices (CCDs) corresponding to a total sensitive area of about 48$\times$72\,mm$^{2}$ \cite{Nelms2002b,Indelicato2006}. 

The Kapton window of the target cell towards the crystal spectrometer had a diameter of 54\,mm. For the geometry as given here, the overall efficiency of the crystal set-up is $\approx5\cdot 10^{-8}$. About 85\% of the reflected intensity is covered by the detector area.

\begin{figure}
% Use the relevant command for your figure-insertion program
% to insert the figure file.
\centering
\includegraphics[width=\columnwidth,clip]{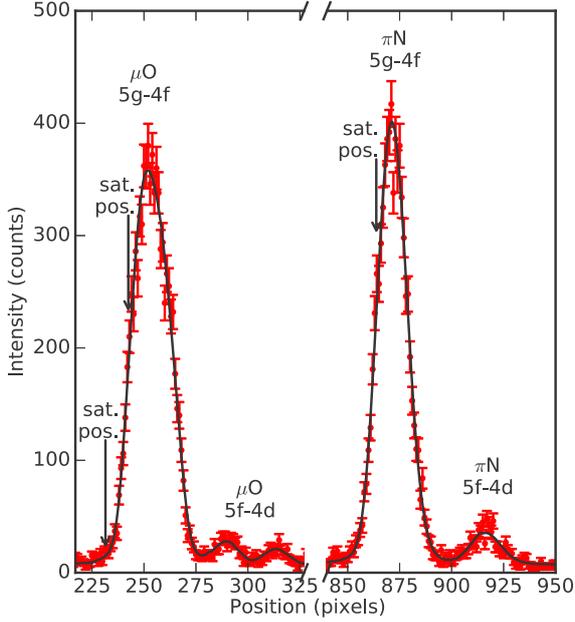}
\caption{Simultaneously measured spectra of $5g-4f$ transitions in muonic oxygen (energy calibration) and pionic nitrogen. The parallel transitions $5f-4d$ and possible additional transitions from the presence of one remaining electron in the K shell are indicated (``sat. pos.''). 
For $\mu$O, the larger width of the $5g-4f$ transition is due to the non-resolved fine structure splitting, which is resolved for the $5f-4d$ transition.}
\label{fig:spectra}       % Give a unique label
\end{figure}

After 5 weeks of data acquisition, about 9000 events were collected of the $(5g-4f)$ transitions for each element.  
The final spectrum is obtained from the projection of the two-dimensional hit pattern on the CCD onto the axis of dispersion. 
The result of the projection is presented in Fig.~\ref{fig:spectra}.

The accuracy of the pion mass measurement essentially depends on the precision of the spectral lines position determination.
This requires both the knowledge of the response function of the spectrometer $F_\text{spectr.}$ and  the control of the effects that the atomic cascade can induce to the line shape via Doppler broadening.
$F_\text{spectr.}$ depends on the characteristic crystal rocking curve but also to aberrations due to the spherical bending of the crystal and the positioning of X-ray source and detector. The imaging properties of the spectrometer were calculated by means of a Monte-Carlo simulation which has been validated by a series of experiments with highly charged ions where natural line widths are negligibly small and high X-ray intensities are available  \cite{Biri2000,Anagnostopoulos2003c,Anagnostopoulos2005,Covita2008,Gotta2016}.

Concerning the atomic cascade processes, the dominant effect on the line width for a molecular target is Coulomb explosion. 
The Doppler broadening and a possible additional broadening due to imperfections of the diffraction crystal are determined from the analysis of a dedicated measurement optimized for pion stops, where in total 60000 events were accumulated in the $\pi$N $5g-4f$ transition. 
From this analysis, we found kinetic energies up to $146 ^{+6}_{-7}$~eV, which corresponds to molecule fragments having about 3.3 elementary charges in agreement with previous results \cite{Siems2000}.
More details of this procedure can be found in Refs.~\cite{Theisen2013,Gotta2015}.

\section{Pion mass evaluation} \label{sec:formulas}

\begin{figure}
% Use the relevant command for your figure-insertion program
% to insert the figure file.
\centering
\includegraphics[width=\columnwidth,clip]{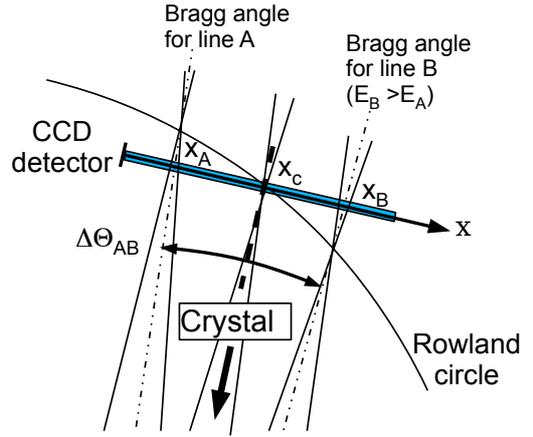}
\caption{Scheme of the X-ray detector position with respect to the Rowland circle (the focus location of the spectrometer) and two diffracted lines $A$ and $B$ with different energy $E_B>E_A$. The (small) displacements of the two lines from their individual focal conditions is taken into account in the calculation of the spectrometer response by means of a Monte-Carlo ray tracing calculation.}
\label{fig:two-lines-spectrum}       % Give a unique label
\end{figure}

The pion mass is deduced from the measurement of the $5g-4f$ $\pi$N transition energy using the $5g-4f$ $\mu$O transition as reference.
The benefits of the choice of this reference are multiple: (i) the reference energy can be calculated very precisely due to the high accuracy of the muon mass value \cite{PDG2014} and (ii) the two lines have almost coinciding energies, {\it i.\,e.} they can be simultaneously measured in the same order of reflection, which reduces drastically the effect of many systematic corrections.

We present here in detail the formulas for deducing the pion mass from the position difference between the spectral line of a pionic atom transition with unknown energy $E_B$ and a reference spectral line with energy $E_A$ (in the same diffraction order).
The basic formula is the Bragg law that relates the X-ray energies to diffraction angles $\Theta$:
\begin{equation}
E_A = n \frac{h c}{2 d \sin \Theta_A}, \label{eq:Bragg}
\end{equation}
where $h$ and $c$ are the Planck and speed of light constant, $d$ is the lattice spacing between crystal planes of the considered reflection direction with order $n$ (an integer number) and $\Theta_A$ is the Bragg angle.

The energy $E_B$ is related to $E_A$ by 
\begin{equation}
E_B =E_A  \frac{1}{\cos \Delta\Theta_{AB} - \cot \Theta_A \sin \Delta\Theta_{AB}}, \label{eq:deltatheta}
\end{equation}
where $\Delta\Theta_{AB} = \Theta_A - \Theta_B$ is the difference between the two Bragg angles, where a small corrections is applied because of the slightly different index of refraction.
$\Delta\Theta_{AB}$ is deduced from the spectral line positions $x_A$ and $x_B$ on the detector plane (see Fig.~\ref{fig:two-lines-spectrum}), which is positioned at a distance $D$ from the crystal.
We have
\begin{equation}
\Delta\Theta_{AB} = \arctan \left( \frac {x_B - x_c} D \right) -  \arctan \left( \frac {x_A - x_c} D \right), \label{eq:deltax}
\end{equation}
where $x_c$ is the position corresponding to the center of the detector.

From Eq.~\eqref{eq:deltatheta} we see that $\cot \Theta_A$ is the only quantity where the values of $h c$ and $d$ play a role according to Eq.\,\eqref{eq:Bragg}.
% with 
%\begin{equation}
%\sin \Theta_A = n \frac{h c}{2 d E_A}.
%\end{equation}
%Rewriting Eq.\,\ref{eq:deltatheta} in terms of $\sin \Theta_A$ we obtain from experiment for the unknown energy 
%\begin{equation}
%E_B =  E_A  \frac{1}{\cos \Delta\Theta_{AB} - \frac {\sqrt{1-\sin^2 \Theta_A}}{\sin \Theta_A} \sin \Delta\Theta_{AB}}. \label{eq:energy}
%\end{equation}

The pion mass $m$ is deduced from the relation of the transition energy $E_B$ 
\begin{equation}
E_B =  \mu_{\pi\mathrm{N}}\,c^2 \frac{(Z \alpha)^2}{2}\left( \frac 1 {n_f^2} - \frac 1 {n_i^2}\right) +\mathcal{O}\left[(Z \alpha)^4 \right]
\end{equation}
to the reduced mass of the pion-nuclear system \begin{equation}
\mu_{\pi\mathrm{N}} = \cfrac{m_\pi}{1+ \cfrac{m_\pi}{M}}, \label{eq:redmass}
\end{equation}
with $M$ being the nuclear mass of nitrogen, $n_i$ and $n_f$ the quantum number of the initial and final state of the transition and $\alpha$ the fine structure constant.
% and the theoretical calculations that provides the dependency between the reduced mass of the system $\mu_{\pi\mathrm{N}}$ and $E_B$
The relationship between $E_B$ and $\mu_{\pi\mathrm{N}}$ contains corrections, not shown explicitly here, that include relativistic effects (additional terms due to the Klein-Gordon equation in our cases), quantum electrodynamics effects (vacuum polarization, self-energy, \dots), recoil corrections, etc. and are summarized here by the term $\mathcal{O}\left[(Z \alpha)^4 \right]$.

%The transition energy $E_B$ is computed by the MCDF program \cite{Desclaux2003,Santos2005,Trassinelli2007}. When considering the charged pion mass value from the Particle Data Group $m_\pi^{PDG}$ \cite{PDG2014} one obtains the expected transition energy $E_B^{PDG}$ to be compared with the experimental value.

For the limited range of pion-mass values, we can assume a linear dependency between $E_B$ and $\mu_{\pi\mathrm{N}}$. 
The proportionality factor is calculated by the MCDF program \cite{Desclaux2003,Santos2005,Trassinelli2007} considering the  charged pion mass value from the Particle Data Group $m_\pi^{PDG}$ \cite{PDG2014} and the corresponding computed transition energy $E_B^{PDG}$. 
In this approximation, the value of $m_\pi$ is given by
\begin{equation}
m = \cfrac{\mu_{\pi\mathrm{N}}^{PDG} \cfrac {E_B}{E_B^{PDG}}}{1 - \cfrac{\mu_{\pi\mathrm{N}}^{PDG}}{M}\cfrac{E_B}{E_B^{PDG}}}~. \label{eq:mass}
\end{equation}

From the line positions and the evaluation of the systematic effects and uncertainties, a pion mass value of  (139.57077\,$\pm$\,0.00018)\,MeV/c$^{2}$ ($\pm 1.3$~ppm) is determined. 
Additional details of the evaluation and of the estimation of the systematic uncertainties are given in Ref.~\citenum{Trassinelli2016b}.
In the next section we specifically discuss the effect of possibly remaining electrons.

\section{Determination of remaining electrons} \label{sec:satellite}

The possible presence of remaining electron(s) in the exotic  atom may induce an important systematic energy shift of the X-ray energies and, consequently, of the pion mass. 
One (or two) remaining electron(s) in the $K$ shell in pionic nitrogen can generate satellite lines having energies 0.45~eV (0.81~eV) lower than the main transition $5g-4f$ \cite{Desclaux2003,Santos2005}.
Such weak satellite lines cannot be resolved from the main transitions, in particular, as they are expected to be of very low intensity (see Fig.~\ref{fig:spectra}). 

As discussed in Sec.~\ref{sec:cascade}, for light pionic atoms  production in a gaseous target of moderate density, the possibility that electrons are present is small when X-ray emission starts. 
However, it cannot be excluded beforehand that small fractions of pions or muons may arrive at the $5g$ level by $\Delta n\,\gg\,1$ transitions from low angular momentum states immediately after capture.
%The presence of remaining electrons is however directly investigated via statistical tests of the present experimental data. 

To estimate the probability for the occurrence of satellite lines we use the evaluation of the Bayes factor \cite{Jeffreys,Sivia,Kass1995,Gordon2007} relative to two hypothesis: the presence or not of satellite lines.
The first hypothesis is associated to the model $M_0$ of the spectra without additional satellite lines.
In the model $M_1$ associated to the second hypothesis, we consider the presence of additional satellite lines with fixed positions with respect to the main transition but keeping the intensity as a free parameter (two lines in the case of muonic oxygen, one per each fine structure main component). 

For each model we sample the likelihood function for different values of parameters to evaluate the Bayesian evidence \cite{Jaynes,Sivia,Trotta2008}, which corresponds to the integral over the parameter space of the likelihood function times the prior probability distributions of the parameters.
The Bayes factor $B_{01}$, the ratio of the Bayesian evidence of the two models, is then calculated as well as the relative probability of the two models.
The evidence is calculated using an homemade code based on the nested sampling algorithm developed by John Skilling in 2004 \cite{Sivia,Skilling2006} (see also Refs.~\citenum{Mukherjee2006,Feroz2009,Veitch2010,Theisen2013} for more details on the calculation method).

An important feature of the evidence calculation is, that  contrary to maximum likelihood and minimum $\chi^2$ methods which provide for each parameter only the most probable value  and the standard deviation, a probability distribution is established for each model parameter (as well as joint probability distributions).

\begin{table}
\centering
\caption{Bayes factor (in logarithmic scale) and the average $P_0$ of the model $M_0$  from the analysis of the different spectra. We also present the possible range of $P_0$ tanking into account the uncertainty of $\ln B_{01}$.}
\label{tab}       % Give a unique label
% For LaTeX tables you can use
\begin{tabular}{p{1.4cm} r r r r}
\hline
Spectrum & $\ln B_{01}$ & $P_0$ &  $P_0^{min}$ & $P_0^{max}$  \\\hline
high-stat. $\pi$N & $6.6 \pm 1.8$ & 99.98\% & 99.86\%  & 100\%   \\
low-stat. $\mu$O &  $-0.3 \pm 0.4$ & 42.52\% & 32.70\% & 52.98\% \\\hline
\end{tabular}
% Or use
%\vspace*{5cm}  % with the correct table height
\end{table}

\begin{figure}
% Use the relevant command for your figure-insertion program
% to insert the figure file.
\centering
\includegraphics[width=0.8\columnwidth,clip]{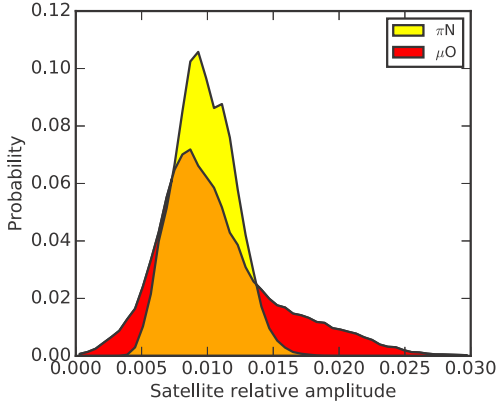}
\caption{Probability distribution of the amplitudes relative to the main line intensity of the possible satellite line due to the presence of one remaining electron in the $K$ shell.}
\label{fig:satellite}       % Give a unique label
\end{figure}

\begin{figure}
% Use the relevant command for your figure-insertion program
% to insert the figure file.
\centering
\includegraphics[width=\columnwidth,clip]{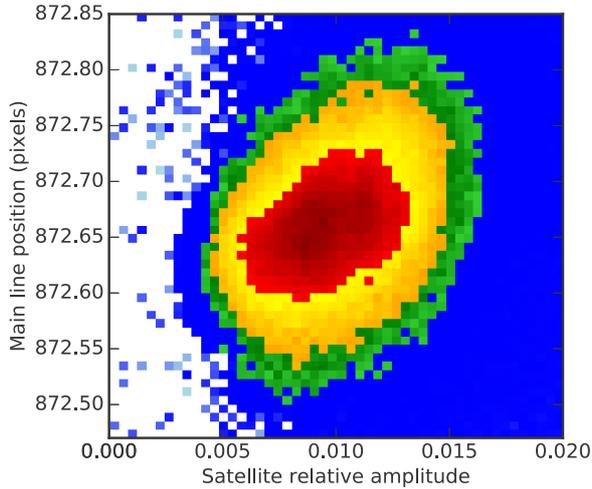}
\caption{Joint probability distribution of the relative satellite amplitude and the position of the main line $5g \to 4f$ in pionic nitrogen. The accumulation on the diagonal shows  the correlation between satellite intensity and main line position. Assuming no satellite line, the position of the main line for this set of data is $872.58 \pm 0.04$ pixels.}
\label{fig:satellite-pos}       % Give a unique label
\end{figure}

For the calculation of $B_{01}$, two data sets were used: the high-statistics spectra of $\pi$N and the low-statistics spectra of $\mu$O (shown in Fig.~\ref{fig:spectra}). The  results are summarized by Figs.~\ref{fig:satellite}, \ref{fig:satellite-pos} and Table~\ref{tab}.
Due to the low statistics, the results from muonic oxygen cannot be used as test against one of the two models  considering the associated uncertainty, because with $\ln B_{01}$ being too close to zero the probability for the two models is similar. For the high-statistics pionic nitrogen spectra, the Bayes factor is significantly different to the unity and the relative probability of the two models can be reliably calculated.
The value $\ln B_{01} = 6.6$ indicates a decisive support for the $M_0$ hypothesis for any Bayes factor scale considered (``decisive'' for Jeffreys scale \cite{Jeffreys}, ``very strong''  for the Kass scale\cite{Kass1995} or ``strong'' for the Gordon-Trotta scale, equivalent to a p-value of about $10^{-5}$ for $M_1$ \cite{Gordon2007}).
Model $M_0$ and $M_1$ relative probabilities are 99.98\% and 0.02\%, respectively.
Though being small, the effect of such a non-zero probability for $M_1$ on the pion mass can be evaluated.
% using the results of the analysis with model with satellites ($M_1$).

When the model $M_1$ is considered, for both sets a satellite amplitude of about 1\% of the main line (see Fig.~\ref{fig:satellite}) is found.
As expected and clearly visible in Fig.~\ref{fig:satellite-pos}, for $\pi$N the satellite amplitude is strongly correlated to the main line position.
We found a shift of the main line of $(\delta x)_1 = 0.08 $~pixels with respect to the case where satellite lines are not taken into account. This is equivalent to about 1~ppm of the pion mass.
This shift cannot be considered in total but has to be weighted by the probability of the two different models using the values of Table~\ref{tab} and where $(\delta x)_0 = 0$.
The expected shift of the main line is then
\begin{equation}
\delta x = 0 \times P_0 + (\delta x)_1 \times P_1 = 1.8 \times 10^{-5}~\text{pixels}
\end{equation}
corresponding to a systematic uncertainty of $^{+0}_{-0.0002}$ ppm. When the Bayes factor accuracy is taken into account  (Table~\ref{tab}) and the worst case is considered, the probability of $M_0$ drops to $P_0^{min}=99.86\%$ and the systematic uncertainty increases to $^{+0}_{-0.0014}$ ppm which is still completely negligible with respect to the statistical error and other systematic effects.

\section{Conclusions}
In conclusion, we present a new measurement of the negatively charged pion mass based on Bragg spectroscopy of pionic nitrogen and muonic oxygen using a gaseous target. 
The use of the low-density target with light atoms prevents  recapture of electrons from neighbouring molecules during the atomic cascade after the formation of the exotic atom.
Consequently, the X rays emitted in the last steps of the cascade stem from a purely hydrogen-like system without systematic effects from unresolved satellite lines due to remaining electrons.
From our high-statistics pionic nitrogen measurement, the probability of presence of satellite peaks has been found to be about 0.02\% confirming the hypothesis on the absence of electrons. Such a contribution introduces a negligible systematic uncertainty on the pion mass of less than one part per billion. 

\hfill

\begin{acknowledgement}
We are grateful to N.\,Dolfus, H.\,Labus, B.\,Leoni and K.-P.\,Wieder for solving numerous technical problems. We thank the PSI staff for providing excellent beam conditions and appreciate the support by the Carl Zeiss AG, Oberkochen, Germany, which fabricated the Bragg crystals. We thank Prof. Dr. E.\,F\"{o}rster and his collaborators at the University of Jena, and A.\,Freund and his group at ESRF, for the help in characterising the crystal material as well as A. Blechmann for a careful study of the CCD performance. We are indebted to PSI for supporting the stay during the run periods (D.\,F.\,A.). This work is part of the PhD thesis of B.\,M. (Universit\'{e} Pierre et Marie Curie, 2001), N.\,N. (University of Leicester, 2002) and M.\,T. (Universit\'{e} Pierre et Marie Curie, 2005).
\end{acknowledgement}

% BibTeX or Biber users please use (the style is already called in the class, ensure that the "woc.bst" style is in your local directory)
\bibliography{Trassinelli_MESON2016}

\begin{thebibliography}{63}

\bibitem{Lattes1947a}
C.M.G. Lattes, G.P.S. Occhialini, C.F. Powell, Nature \textbf{160}, 453 (1947)

\bibitem{Lattes1947b}
C.M.G. Lattes, G.P.S. Occhialini, C.F. Powell, Nature \textbf{160}, 486 (1947)

\bibitem{Gardner1948}
E.~Gardner, C.M.G. Lattes, Science \textbf{107}, 270 (1948)

\bibitem{Barkas1951}
W.H. Barkas, F.M. Smith, E.~Gardner, Phys. Rev. \textbf{82}, 102 (1951)

\bibitem{Smith1953}
F.M. Smith, W.~Birnbaum, W.H. Barkas, Phys. Rev. \textbf{91}, 765 (1953)

\bibitem{Barkas1956}
W.H. Barkas, W.~Birnbaum, F.M. Smith, Phys. Rev. \textbf{101}, 778 (1956)

\bibitem{Crowe1954}
K.M. Crowe, R.H. Phillips, Phys. Rev. \textbf{96}, 470 (1954)

\bibitem{Stearns1954}
M.~Stearns, M.B. Stearns, S.~DeBenedetti, L.~Leipuner, Phys. Rev. \textbf{95},
  1353 (1954)

\bibitem{Shafer1967}
R.E. Shafer, Phys. Rev. \textbf{163}, 1451 (1967)

\bibitem{Marushenko1976}
V.N. Marushenko, A.F. Mezentsev, A.A. Petrunin, S.G. Skornyakov, A.I. Smirnov,
  JETP Lett. \textbf{23}, 72 (1976)

\bibitem{Carter1976}
A.L. Carter, M.S. Dixit, M.K. Sundaresan, J.S. Wadden, F.J. Watson, C.K.
  Hargrove, E.P. Hincks, R.J. McKee, H.~Mes, H.L. Anderson et~al., Phys. Rev.
  Lett. \textbf{37}, 1380 (1976)

\bibitem{Lu1980}
D.C. Lu, D.~Delker, G.~Dugan, C.S. Wu, A.J. Caffrey, Y.T. Cheng, Y.K. Lee,
  Phys. Rev. Lett. \textbf{45}, 1066 (1980)

\bibitem{Jeckelmann1986}
B.~Jeckelmann, T.~Nakada, W.~Beer, G.~de~Chambrier, O.~Elsenhans, K.L.
  Giovanetti, P.F.A. Goudsmit, H.J. Leisi, A.~Rüetschi, O.~Piller et~al.,
  Phys. Rev. Lett. \textbf{56}, 1444 (1986)

\bibitem{Jeckelmann1986b}
B.~Jeckelmann, W.~Beer, G.~De~Chambrier, O.~Elsenhans, K.L. Giovanetti, P.F.A.
  Goudsmit, H.J. Leisi, T.~Nakada, O.~Piller, A.~Rüetschi et~al., Nucl. Phys.
  A \textbf{457}, 709 (1986)

\bibitem{Jeckelmann1994}
B.~Jeckelmann, P.F.A. Goudsmit, H.J. Leisi, Phys. Lett. B \textbf{335}, 326
  (1994)

\bibitem{Lenz1998}
S.~Lenz, G.~Borchert, H.~Gorke, D.~Gotta, T.~Siems, D.F. Anagnostopoulos,
  M.~Augsburger, D.~Chatellard, J.P. Egger, D.~Belmiloud et~al., Phys. Lett. B
  \textbf{416}, 50 (1998)

\bibitem{Abela1984}
R.~Abela, M.~Daum, G.H. Eaton, R.~Frosch, B.~Jost, P.R. Kettle, E.~Steiner,
  Phys. Lett. B \textbf{146}, 431 (1984)

\bibitem{Daum1991}
M.~Daum, R.~Frosch, D.~Heter, M.~Janousch, P.R. Kettle, Phys. Lett. B
  \textbf{265}, 425 (1991)

\bibitem{Assamagan1996}
K.~Assamagan, C.~Brnnimann, H.F. M.~Daum, R.~Frosch, P.~Gheno, R.~Horisberger,
  M.~Janousch, P.R. Kettle, T.~Spirig, C.~Wigger, Phys. Rev. D \textbf{53},
  6065 (1996)

\bibitem{Gotta1999}
D.~Gotta, D.F. Anagnostopoulos, M.~Augsburger, G.~Borchert, C.~Castelli,
  D.~Chatellard, J.P. Egger, P.~El-Khoury, H.~Gorke, P.~Hauser et~al., Nucl.
  Phys. A \textbf{660}, 283 (1999)

\bibitem{PDG2014}
K.A. Olive, G.~Particle~Data, Chinese Phys. B \textbf{38}, 090001 (2014)

\bibitem{Trassinelli2016b}
M.~Trassinelli, D.F. Anagnostopoulos, G.~Borchert, A.~Dax, J.P. Egger,
  D.~Gotta, M.~Hennebach, P.~Indelicato, Y.W. Liu, B.~Manil et~al., Phys. Lett.
  B \textbf{759}, 583 (2016)

\bibitem{Simons1988}
L.M. Simons, Phys. Scripta \textbf{T22}, 90 (1988)

\bibitem{Simons1993}
L.M. Simons, Hyperfine Interact. \textbf{81}, 253 (1993)

\bibitem{Cohen2004}
J.S. Cohen, Rep. Prog. Phys. \textbf{67}, 1769 (2004)

\bibitem{Burbidge1953}
G.R. Burbidge, A.H. de~Borde, Phys. Rev. \textbf{89}, 189 (1953)

\bibitem{Leon1962}
M.~Leon, H.A. Bethe, Phys. Rev. \textbf{127}, 636 (1962)

\bibitem{Vogel1980}
P.~Vogel, Phys. Rev. A \textbf{22}, 1600 (1980)

\bibitem{Bacher1985}
R.~Bacher, D.~Gotta, L.M. Simons, J.~Missimer, N.C. Mukhopadhyay, Phys. Rev.
  Lett. \textbf{54}, 2087 (1985)

\bibitem{Bacher1989}
R.~Bacher, P.~Blüm, D.~Gotta, K.~Heitlinger, M.~Schneider, J.~Missimer, L.M.
  Simons, Phys. Rev. A \textbf{39}, 1610 (1989)

\bibitem{Kirch1999}
K.~Kirch, D.~Abbott, B.~Bach, P.~Hauser, P.~Indelicato, F.~Kottmann,
  J.~Missimer, P.~Patte, R.T. Siegel, L.M. Simons et~al., Phys. Rev. A
  \textbf{59}, 3375 (1999)

\bibitem{Siems2000}
T.~Siems, D.F. Anagnostopoulos, G.~Borchert, D.~Gotta, P.~Hauser, K.~Kirch,
  L.~Simons, P.~El-Khoury, P.~Indelicato, M.~Augsburger et~al., Phys. Rev.
  Lett. \textbf{84}, 4573 (2000)

\bibitem{Jensen2002a}
T.S. Jensen, V.E. Markushin, Eur. Phys. J. D \textbf{19}, 165 (2002)

\bibitem{Jensen2002b}
T.S. Jensen, V.E. Markushin, Eur. Phys. J. D \textbf{21}, 261 (2002)

\bibitem{Kilic2004}
S.~Kilic, J.P. Karr, L.~Hilico, Phys. Rev. A \textbf{70}, 042506 (2004)

\bibitem{Pomerantsev2006}
V.N. Pomerantsev, V.P. Popov, Phys. Rev. A \textbf{73}, 040501 (2006)

\bibitem{Jensen2007}
T.~Jensen, V.~Popov, V.~Pomerantsev, arXiv preprint arXiv:0712.3010  (2007)

\bibitem{Popov2007}
V.~Popov, V.~Pomerantsev, arXiv preprint arXiv:0712.3111  (2007)

\bibitem{Johann1931}
H.H. Johann, Z. f. Physik \textbf{69}, 185 (1931)

\bibitem{Gotta2004}
D.~Gotta, Progress in Particle and Nuclear Physics \textbf{52}, 133 (2004)

\bibitem{Gotta2016}
D.E. Gotta, L.M. Simons, Spectrochim. Acta, Part B \textbf{120}, 9 (2016)

\bibitem{Eggs1965}
J.~Eggs, K.~Ulmer, Zf. angew. Phys. \textbf{20}, 118 (1965)

\bibitem{Nelms2002b}
N.~Nelms, D.F. Anagnostopoulos, O.~Ayranov, G.~Borchert, J.P. Egger, D.~Gotta,
  M.~Hennebach, P.~Indelicato, B.~Leoni, Y.W. Liu et~al., Nucl. Instrum.
  Methods A \textbf{484}, 419 (2002)

\bibitem{Indelicato2006}
P.~Indelicato, E.O. {Le~Bigot}, M.~Trassinelli, D.~Gotta, M.~Hennebach,
  N.~Nelms, C.~David, L.M. Simons, Rev. Sci. Instrum. \textbf{77}, 043107 (~10)
  (2006)

\bibitem{Biri2000}
S.~Biri, L.~Simons, D.~Hitz, Rev. Sci. Instrum. \textbf{71}, 1116 (2000)

\bibitem{Anagnostopoulos2003c}
D.F. Anagnostopoulos, S.~Biri, V.~Boisbourdain, M.~Demeter, G.~Borchert, J.P.
  Egger, H.~Fuhrmann, D.~Gotta, A.~Gruber, M.~Hennebach et~al., Nucl. Instrum.
  Methods B \textbf{205}, 9 (2003)

\bibitem{Anagnostopoulos2005}
D.F. Anagnostopoulos, S.~Biri, H.~Fuhrmann, D.~Gotta, A.~Gruber, P.~Indelicato,
  B.~Leoni, L.M. Simons, L.~Stingelin, A.~Wasser et~al., Nucl. Instrum. Methods
  A \textbf{545}, 217 (2005), \texttt{physics/0408081}

\bibitem{Covita2008}
D.S. Covita, Ph.D. thesis, University of Coimbra (2008)

\bibitem{Theisen2013}
M.~Theisen, Diplomarbeit, Fakultät für Mathematik, Informatik uns
  Naturwisssenschaften der RWTH Aachen (2013)

\bibitem{Gotta2015}
D.~Gotta, F.D. Amaro, D.F. Anagnostopoulos, P.~Bühler, H.~Gorke, D.S. Covita,
  H.~Fuhrmann, A.~Gruber, M.~Hennebach, A.~Hirtl et~al., Hyperfine Interact.
  \textbf{234}, 105 (2015)

\bibitem{Desclaux2003}
J.~Desclaux, J.~Dolbeault, M.~Esteban, P.~Indelicato, E.~Séré, in
  \emph{Computational Chemistry}, edited by C.~Le~Bris, M.~De~Franceschi
  (Elsevier, Amsterdam, 2003), Vol.~X of \emph{Handbook of Numerical Analysis},
  p. 1032

\bibitem{Santos2005}
J.P. Santos, F.~Parente, S.~Boucard, P.~Indelicato, J.P. Desclaux, Phys. Rev. A
  \textbf{71}, 032501 (~8) (2005)

\bibitem{Trassinelli2007}
M.~Trassinelli, P.~Indelicato, Phys. Rev. A \textbf{76}, 012510 (2007)

\bibitem{Jeffreys}
H.~Jeffreys, \emph{Theory of Probability}, 3rd~edn. (Oxford University Press,
  Oxford, U.K., 1961)

\bibitem{Sivia}
D.S. Sivia, J.~Skilling, \emph{Data analysis: a Bayesian tutorial}, 2nd~edn.
  (Oxford University Press, 2006)

\bibitem{Kass1995}
R.E. Kass, A.E. Raftery, J. Acoust. Soc. America \textbf{90}, 773 (1995)

\bibitem{Gordon2007}
C.~Gordon, R.~Trotta, Mon. Not. R. Astron. Soc. \textbf{382}, 1859 (2007)

\bibitem{Jaynes}
E.~Jaynes, G.~Bretthorst, \emph{Probability Theory: The Logic of Science}
  (Cambridge University Press, 2003), ISBN 9781139435161

\bibitem{Trotta2008}
R.~Trotta, Contemp. Phys. \textbf{49}, 71 (2008)

\bibitem{Skilling2006}
J.~Skilling, Bayesian Anal. \textbf{1}, 833 (2006)

\bibitem{Mukherjee2006}
P.~Mukherjee, D.~Parkinson, A.R. Liddle, Astrophys. J. Lett. \textbf{638}, L51
  (2006)

\bibitem{Feroz2009}
F.~Feroz, M.P. Hobson, M.~Bridges, Mon. Not. R. Astron. Soc. \textbf{398}, 1601
  (2009)

\bibitem{Veitch2010}
J.~Veitch, A.~Vecchio, Phys. Rev. D \textbf{81}, 062003 (2010)

\end{thebibliography}

\end{document}